\documentclass[preprint,12pt]{elsarticle}

\usepackage{amssymb,amsthm,latexsym,amsmath,amsfonts}

\journal{Physica D}

\newtheorem{thm}{Theorem}
\newtheorem{cor}{Corollary}
\newtheorem{lm}{Lemma}
\newtheorem{prop}{Proposition}

\begin{document}

\begin{frontmatter}

\title{Topological Quantification of the ``Anemone'' (Branching) Solar Flares}

\author[HSE_NN]{Evgeny~V.~Zhuzhoma}
\ead{zhuzhoma@mail.ru}

\author[HSE_NN]{Vladislav~S.~Medvedev}
\ead{medvedev-1942@mail.ru}

\author[SAI,IKI]{Yurii~V.~Dumin\corref{corauthor}}
\ead{dumin@yahoo.com}

\author[SAI]{Boris~V.~Somov}
\ead{somov-boris@mail.ru}

\address[HSE_NN]{HSE University,
Faculty of Informatics, Mathematics, and Computer Science\\
(Nizhny Novgorod Campus),\\
25/12 Bolshaya Pecherskaya ulitsa, 603155 Nizhny Novgorod, Russia}

\address[SAI]{Lomonosov Moscow State University~(MSU),
Sternberg Astronomical Institute~(GAISh),\\
13 Universitetskii prospekt, 119234 Moscow, Russia}

\address[IKI]{Space Research Institute~(IKI)
of Russian Academy of Sciences,\\
84/32 Profsoyuznaya ulitsa, 117997 Moscow, Russia}

\cortext[corauthor]{Corresponding author.}

\begin{abstract}
The so-called ``anemone'' solar flares are an interesting type
of the space plasma phenomena, where multiple null points of the magnetic
field are connected with each other and with the magnetic sources by
the separators, thereby producing the complex branching configurations.
Here, using the methods of dynamical systems and Morse--Smale theory,
we derive a few universal topological relations between the numbers of
the null points and sources of various kinds with arbitrary arrangement
in the above-mentioned structures.
Such relations can be a valuable tool both for a quantification of
the already-observed anemone flares and for a prediction of the new ones
in complex magnetic configurations.
\end{abstract}


\end{frontmatter}

\section{Introduction}
\label{sec:Intro}

The solar flares are among the most energetic phenomena in the Solar System,
substantially affecting our space environment.
They are commonly assumed to be produced by the so-called magnetic
reconnection, when the magnetic field lines break and then merge with each
other in a new configuration, while the excessive energy is released in
the form of the heated plasmas and accelerated particles~\cite{Priest_00,%
Somov_13}.

From the geometric point of view, the flares are usually formed by the sets of
magnetic arcades, rooted at the solar surface (or the so-called photosphere)
and extending up to the upper layers (the solar corona).
These arcs can be immediately observed in the hard ultraviolet and X-rays,
while their footpoints are usually observable in the visible light as
two approximately parallel ribbons~\cite{Huang_18}.
In some cases, the magnetic arcades can intersect each other, forming more
complex spatial configurations~\cite{Nikulin_16}, but their topology remains
quite trivial.

\begin{figure}
\center{\includegraphics[width=12cm]{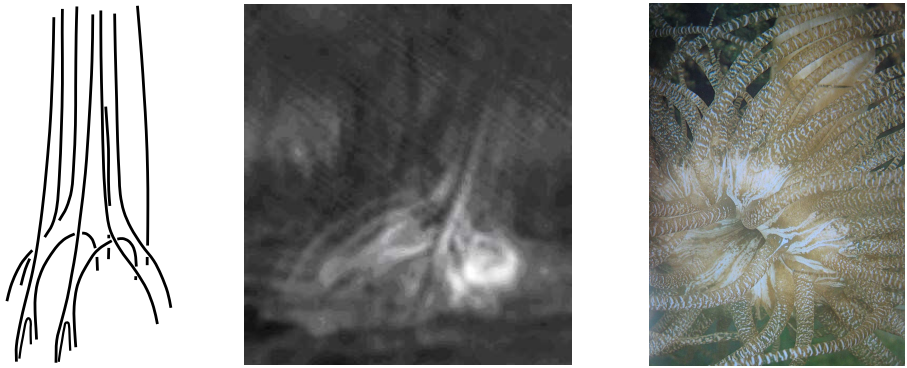}}
\caption{\label{fig:observ_data}
The hypothesized structure of magnetic field lines in the anemone
flare~\cite{Shibata_07} (left panel, reprinted with permission from
the American Association for the Advancement of Science {\copyright}2007)
vs.\ the picture taken by the New Solar Telescope in the Big Bear Solar
Observatory~\cite{Zeng_16} (middle panel, reprinted with permission from
the American Astronomical Society {\copyright}2016),
as well as an example of the biological anemone (right panel, courteously
provided by G.A.~Porfir'eva).}
\end{figure}

On the other hand, a much more sophisticated topology can be realized in the
so-called anemone microflares, occurring in the solar chromosphere (i.e.,
a bit above the photosphere).
The first hints to this phenomenon were given by observations of Hinode
satellite~\cite{Shibata_07}.
Namely, a few diverging small-scale luminous ribbons were found in the base
of such flares.
Then, they were qualitatively interpreted as footpoints of the magnetic field
lines experiencing the bifurcations (branching) at some height in the course
of magnetic reconnection (left panel in Fig.~\ref{fig:observ_data}).
A decade later, such bifurcations became directly observable by the New Solar
Telescope in the Big Bear Solar Observatory (California, USA)~\cite{Zeng_16};
a particular example is presented in the middle panel of
Fig.~\ref{fig:observ_data}.
At last, the right panel of this figure illustrates a remarkable similarity
of such flares with the sea anemone, which are well known in biology.

A theoretical interpretation of the above-mentioned phenomenon requires
a consideration of bifurcations of the solar magnetic fields, for example,
in the framework of the potential field model formed by the effective
point-like charges.

\subsection{The Concept of the Effective Magnetic Charges}
\label{sec:Magnetic_charges}

\begin{figure}
\center{\includegraphics[width=14cm]{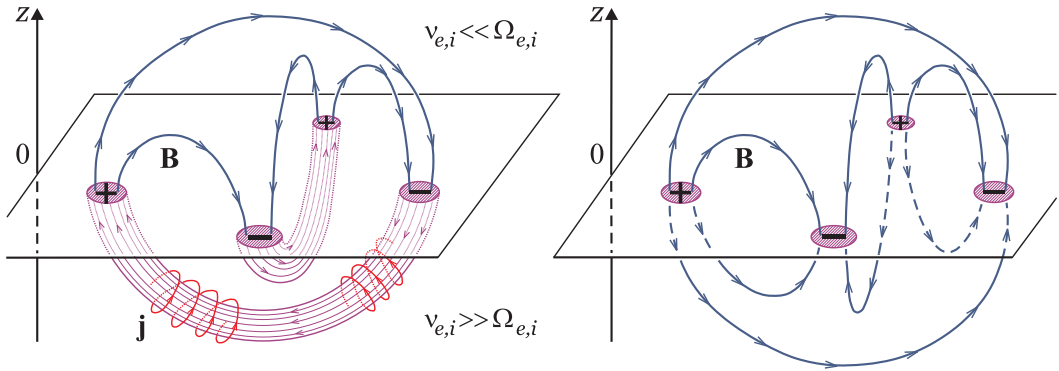}}
\caption{\label{fig:magn_charges}
Magnetic-field sources and sinks in the solar atmosphere, caused by
the open ends of the magnetic-flux tubes under the photosphere (left
panel), and their formal reduction to the magnetic charges (positive and
negative, respectively) in the plane $ z\,=\,0 $ (right panel).}
\end{figure}

Since both the present paper and a number of previous studies on the
topology of solar magnetic fields were substantially based on the idea
of the effective magnetic charges, it it reasonable to explain this
concept in more detail.
Strictly speaking, any magnetic field is non-divergent
($ {\rm div}\,{\bf B} = 0 $).
Therefore, its field lines are closed, and there cannot exist any magnetic
charges (sources and sinks).
However, it is often convenient to introduce the ``effective'' magnetic
charges in the sense illustrated in Fig.~\ref{fig:magn_charges}.

Namely, the electric currents $ {\bf j} $ in the deep layers of the Sun
(where the collisional frequencies of both electrons and ions are much
greater than their gyrofrequencies,
$ {{\nu}_{e,i}}\,{\gg}\,{{\Omega}_{e,i}} $)
form the tubes of the concentrated magnetic flux.
The open ends of such tubes at the surface of photosphere, $ z = 0 $,
serve as the sources and sinks of the magnetic field in the upper layers
of the Sun, which are collisionless,
$ {{\nu}_{e,i}}\,{\ll}\,{{\Omega}_{e,i}} $
(left panel in Fig.~\ref{fig:magn_charges}).
Moreover, in certain circumstances, this magnetic field is approximately
current-free, i.e., potential ($ {\rm rot}\,{\bf B} = 0 $).

Next, one can consider only the upper semispace ($ {z}\,{\ge}\,{0} $) and
formally perform its mirror reflection with respect to the plane
$ {z}\,{=}\,{0} $
(right panel in the same figure).
As a result, we get a symmetric pattern of the magnetic-field lines,
whose sources and sinks (the effective ``magnetic charges'') are located
exactly in the plane $ {z}\,{=}\,{0} $.
Therefore, as was pictorially outlined in paper~\cite{Longcope_96},
``magnetic field enters the corona from the interior of the Sun through
isolated magnetic features on the solar surface.
These features correspond to the tops of submerged magnetic flux tubes,
and coronal field lines often connect one flux tube to another, defining
a pattern of inter-linkage.
Using a model field, in which flux tubes are represented as point magnetic
charges, it is possible to quantify this inter-linkage.''

The potential field generated by the set of charges (monopoles) is similar
to the electrostatic field, and it is quite convenient for the subsequent
mathematical analysis.
Of course, some care must be taken in the interpretation of the corresponding
mathematical results.
For example, if one have found some number of peculiarities of the field
(e.g., null points) beyond the plane $ {z}\,{=}\,{0} $, then only one half
of them will have a real physical meaning (namely, those located in
the upper semispace).
On the other hand, if such peculiarities are localized exactly in the plane
$ {z}\,{=}\,{0} $, then all of them should be treated as physically relevant.

\subsection{Review of the Previous Studies}
\label{sec:Review}

While the term ``topology of magnetic field'' is widely employed in the
literature on solar physics, there were actually a very few papers devoted
to the rigorous topological analysis of the respective magnetic
configurations.
They were usually based on the computer simulations supplemented by some
analytical results from the algebraic topology.
One of the first works of this kind was paper~\cite{Baum_80}, whose authors
analyzed a few particular configurations of the magnetic field produced
by the four magnetic charges (two positive and two negative) with equal
magnitudes.
A much more general analysis of approximately the same situation was
performed in paper~\cite{Gorbachev_88}, where four magnetic charges were
allowed to be arbitrary located in the plane of the photosphere.
Next, employing some theorems of differential geometry and algebraic
topology, the authors established the general criteria for the existence of
null points of various types (both in and out of the plane of charges)
depending on the localization of the charges.
The most interesting finding was that there are such positions of
the magnetic charges when a tiny displacement of one of them results in
the emergence of a new null point and its fast motion over a considerable
distance high above the plane of the sources.
This fact inspired a new mechanism of the magnetic reconnection,
the so-called ``topological trigger''~\cite{Somov_08}; examples of its
practical application to the particular flares can be found, e.g.,
in paper~\cite{Oreshina_12}.

Next, bifurcations of the null points in the systems formed by three and
four unbalanced irregularly-located charges were analyzed in
paper~\cite{Brown_99}.
On the other hand, paper~\cite{Inverarity_99} dealt with a highly-symmetric
configuration: the numerous positive magnetic charges (sources) were
localized in the nodes of a hexagonal network (mimicking the so-called
supergranule convective cells) and a single negative charge (sink) was
placed in the center of this structure.
Then, the authors sought for the null points both in and above the plane of
the charges, as well as studied their emergence and displacement depending on
the magnitude of the central sink and its shift from the center of symmetry.
A further discussion of both symmetric and irregular magnetic-charge
configurations with special emphasis on the emergence of the ``off-plane''
null points was given in paper~\cite{Brown_01}.
Review of application of various topological methods in the solar physics
can be found in paper~\cite{Longcope_05}.

As follows from the above discussion, the previous works treated either
configurations with rather symmetric arrangement or very small number of
the magnetic charges.
On the other hand, in view of the recent interest to the anemone (branching)
solar flares, it would be important to get criteria for the emergence of
null points when the magnetic charges are numerous and located irregularly.
So, it is the aim of the present paper to perform such analysis by using
the Morse--Smale theory of vector fields.%
\footnote{
Yet another application of the Morse--Smale theory to the analysis of magnetic
fields in plasmas can be found in paper~\cite{Grines_15}.}

\section{Summary of the main results}
\label{sec:Summary}

The magnetic charge is called \textit{positive} if the field flux through
an arbitrary small sphere covering the charge is positive.
The \textit{negative} charge is defined by a similar way: the field flux
through an arbitrary small sphere covering such a charge is negative.

The group of charges~$ \mathcal{C}$ is called \textit{positively unbalanced}
if it can be embedded into the ball~$ B $ so that the magnetic field is
directed outwards at its boundary.
The above-specified ball~$ B = B(\mathcal{C}) $ will be called
\textit{a source region of the group}~$\mathcal{C}$.
The \textit{negatively unbalanced} group of charges is defined by a similar
way, and it is associated with \textit{a sink region of the group}.

\begin{figure}
\center{\includegraphics[width=10cm]{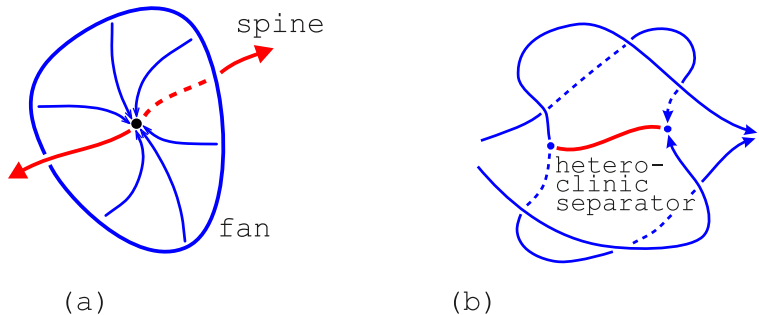}}
\caption{\label{fig:spine-fan}
Structure of the null point~(a) and the heteroclinic separator~(b).}
\end{figure}

The idealized magnetic charge corresponds to a point-like singularity
in the vector field; the positive charge being considered as a source and
the negative charge as a sink of the field. The point~$ p_0 $ of
the magnetic field~$ {\bf B} $ is called \textit{the null point}
if $ {\bf B}(p_0) = 0 $.
The eigenvalues $ {\lambda}_1 $, $ {\lambda}_2 $ and $ {\lambda}_3 $ in
the null point are typically nonzero and satisfy the equality
$ {\lambda}_1 + {\lambda}_2 + {\lambda}_3 = 0 $, because
$ {\nabla}{\cdot}{\bf B} = 0 $.
Consequently, from the viewpoint of the theory of dynamical systems,
the null point is a conservative saddle, possessing one 1D and one
2D separatrices; see figure~\ref{fig:spine-fan}(a).%
\footnote{
The 1D separatrix is sometimes called \textit{the spine}; and 2D separatrix,
\textit{the fan}~\cite{Priest_00}.}
If the magnetic field line on the 1D separatrix is directed from the null
point, then all field lines on the 2D separatrix surface are directed to
the null point; and vice versa.

The following two cases are possible for a typical null point~$ p_0 $
(up to redefinition of the eigenvalues):\\
(1) $ {\lambda}_1 > 0 $, \quad $ {\lambda}_2 , {\lambda}_3 < 0 $;\\
(2) $ {\lambda}_1 < 0 $, \quad $ {\lambda}_2 , {\lambda}_3 > 0 $.

In the first case, the null point~$ p_0 $ is called \textit{positive},
because $ {\lambda}_1{\cdot}{\lambda}_2{\cdot}{\lambda}_3~>~0 $.
From the viewpoint of the theory of dynamical systems, the positive null
point represents a saddle with Morse index equal to 1 and topological index
equal to $ -1 $.
Such saddle has a 1D unstable separatrix and a 2D stable separatrix.
In the second case, the point~$ p_0 $ is called~\textit{negative}, because
$ {\lambda}_1{\cdot}{\lambda}_2{\cdot}{\lambda}_3~<~0 $.
Such null point is a saddle with Morse index equal to 2 and topological
index equal to $ +1 $.
This saddle has a 1D stable separatrix and 2D unstable separatrix;
see Fig.~\ref{fig:spine-fan}(a).

Following the standard terminology~\cite{Priest_00},
the magnetic field line connecting two null points will be called
\textit{a separator}.The separator is called \textit{heteroclinic} if it
represents a transversal intersection of two separatrix surfaces;
see Fig.~\ref{fig:spine-fan}(b).
Topological structure of the magnetic field is determined by the number and
types of the null points, by the location of spines and fans with respect to
each other, and by the lines of transversal intersection of the fan surfaces,
i.e., the heteroclinic separators.
Later on, for simplicity, a separator means the heteroclinic separator.

\begin{thm}
\label{thm:at-least-many-nul-points}
Let a positively unbalanced group~$ \mathcal{C} $ contain
$ l \geq 2 $~positive charges (and arbitrary number of negative charges).
Then there exist at least $ l-1 $ negative null points in the source
region~$ B(\mathcal{C}) $ of this group.
If the group~$ \mathcal{C} $ consists of $ l \geq 2 $~positive charges
and there are exactly $ l-1 $ null points in~$ B(\mathcal{C}) $, then all
these null points are negative and there are no separators
in~$ B(\mathcal{C}) $.
Moreover, the magnetic field in region~$ B(\mathcal{C}) $ possesses
a uniquely defined, up to the topological equivalence, structure.
\end{thm}

The next consequences follow from this theorem.

\begin{cor}
\label{cor1:from-thm-at-least-many-nul-points}
Let a negatively unbalanced group~$ \mathcal{C} $ contain
$ k \geq 2 $~negative charges.
Then there exist at least $ k-1 $ positive null points in the sink
region~$ B(\mathcal{C}) $ of this group.
If the group~$ \mathcal{C} $ consists of $ k \geq 2 $~negative charges
and there are exactly $ k-1 $ null points in~$ B(\mathcal{C}) $, then all
these null points are positive and there are no separators
in~$ B(\mathcal{C}) $.
\end{cor}

\begin{cor}
\label{cor2:from-thm-at-least-many-nul-points}
Let a positively unbalanced group~$ \mathcal{C} $ contain one (dominant)
positive charge and $ k \geq 1 $~negative charges.
Then there exist at least $ k $~positive null points in the source
region~$ B(\mathcal{C}) $ of this group.
If $B(\mathcal{C})$ contains exactly $ k $~null points, then all these
points are positive and there are no separators in~$B(\mathcal{C})$.
Moreover, the magnetic field in region~$ B(\mathcal{C}) $ possesses
a uniquely defined, up to the topological equivalence, structure.
\end{cor}

\begin{cor}
\label{cor3:from-thm-at-least-many-nul-points}
Let a positively unbalanced group~$ \mathcal{C} $ contains
$ l \geq 2 $ positive and $ k \geq 1 $ negative charges.
Then there exist at least $ l-1 $~negative null points and
at least $ k $~positive null points in the source
region~$ B(\mathcal{C}) $ of this group.
\end{cor}

At the minimal numbers of both the positive and negative null points (which
are determined by Corollary~\ref{cor3:from-thm-at-least-many-nul-points}),
the separators can be absent.
Nevertheless, as follows from the subsequent theorem, as soon as
at least one ``excessive'' null point appears, at least one separator
is inevitably formed.
The type of the excessive null point is of no importance: it can be either
positive or negative.
For the sake of definiteness, we shall consider the case when the excessive
point is negative.

\begin{thm}
\label{thm:at-least-one-separator}
Let a positively unbalanced group~$ \mathcal{C} $ contain
$ l \geq 2 $ positive and $ k \geq 0 $ negative charges.
If $ B(\mathcal{C}) $ contains exactly $ l $~negative null points, then
there is at least one separator in~$B(\mathcal{C})$.
\end{thm}

The last theorem demonstrates a particular scenario of emergence of
a negative null point, when there is a family of separators whose number
is equal to the number of negative null points (and it is equal also to
the number of positive charges).

\begin{thm}\label{thm:birth-many-separators}
Let a positively unbalanced group~$ \mathcal{C} $ contain $ l \geq 2 $
positive charges and $ l-1 $ negative null points, such that this group is
defined by the vector field~$\vec{F}_0$.
Then, there is a continuous family of vector fields~$\vec{F}_{t}$,
$ 0 \leq t \leq 1 $, such that the vector field~$ \vec{F}_1 $
has~$ l $ positive charges, $ l $~separators, and one negative null point.
\end{thm}

\section{Auxiliary information}
\label{sec:Auxiliary}

Let $ f^t $~be a flow induced by the vector field~$ {\bf V} $ on
the 3D sphere~$ \mathbb{S}^3 $.
We shall assume that $ f^t $ has no periodic trajectories.
Let $ Fix~(f^t) $ designate a set of equilibrium states of the flow~$ f^t $.
For~$ p \in Fix~(f^t) $, let $ W^s(p) $ be a set of trajectories
approaching~$ p $ at infinitely increasing time.%
\footnote{
From here on, the independent variable~$ t $ will be called time,
as it is commonly accepted in the theory of dynamical systems.
However, it should be kept in mind that from the physical point of view
this is actually a variable parametrizing the length of the magnetic field
line.}
In particular, if $ p $~is a saddle, then $ W^s(p) \setminus \{p\} $ is
a stable separatrix of the saddle~$ p $.
The set~$ W^s(p) $ is called \textit{the stable manifold} of the point~$ p $.
Similarly, let $ W^u(p) $ denote a set of trajectories approaching~$ p $
at infinitely decreasing time.
In particular, if $ p $~is a saddle, then $ W^u(p) \setminus \{p\} $~is
an unstable separatrix of the saddle~$ p $.
The set~$ W^u(p) $ is called \textit{the unstable manifold} of the point~$p$.
The flow~$ f^t $ is called \textit{the Morse--Smale flow} if all its
equilibrium states are hyperbolic, their stable and unstable manifolds
intersect each other transversally, and the limit set for any trajectory
belongs to~$ Fix~(f^t) $.
The corresponding vector field~$ {\bf V} $ is called
\textit{the Morse--Smale vector field}~\cite{Smale_67}.

Let $ {\bf v}_{\rm sourse} $ denote the vector field in
ball~$ \mathbb{B}^3 $ directed outwards at the ball
boundary~$ \partial\mathbb{B}^3 = S^2 $ and possessing exactly one
source inside the ball.
Let us assume that
$ {\bf v}_{\rm sink} = -{\bf v}_{\rm sourse} $.
Obviously, the vector field~$ {\bf v}_{\rm sink} $ possesses
exactly one sink inside the ball~$ \mathbb{B}^3 $, and
$ {\bf v}_{\rm sink}$~is directed inwards the ball~$ \mathbb{B}^3 $
at the boundary~$ S^2 $.

Let $ B(\mathcal{C}) $~be the source region of the group of
charges~$ \mathcal{C} $.
Let $ {\bf v}(\mathcal{C}) $ denote the magnetic field
in~$ B(\mathcal{C}) $ created by the group~$ \mathcal{C} $.
We remind that the vector field~$ {\bf v}(\mathcal{C}) $ is directed
outwards at the boundary~$ \partial B(\mathcal{C}) $ of
the ball~$ B(\mathcal{C}) $.
If boundaries~$ \partial\mathbb{B}^3 = S^2 $ and $ \partial B(\mathcal{C}) $
of the balls~$ \mathbb{B}^3 $ and~$ B(\mathcal{C}) $, respectively, are
identical to each other, then we get a 3D sphere~$ \mathbb{S}^3 $.
The field~$ {\bf v}_{\rm sink} $ near
the boundary~$ \partial\mathbb{B}^3 $ can be corrected so that
the fields~$ {\bf v}_{\rm sink} $ and~$ {\bf v}(\mathcal{C}) $
form a smooth Morse--Smale vector field at~$ \mathbb{S}^3 $, which will be
denoted by~$ {\bf V}(\mathcal{C}) $.
Obviously, a global topological structure of
the field~$ {\bf v}_{\rm sink} $ can be preserved after such
transformation.
Then the equilibrium states of~$ {\bf V}(\mathcal{C}) $ will
represent a union of the equilibrium states for
the field~$ {\bf v}(\mathcal{C}) $ and the sink
field~$ {\bf v}_{\rm sink} $.
The vector field~$ {\bf V}(\mathcal{C}) $ will be called
\textit{a continuation of the field}~$ {\bf v}(\mathcal{C})$ by
the group~$ \mathcal{C} $ to the 3D sphere~$ \mathbb{S}^3 $.

\begin{lm}
\label{lm:eiler-poincare-formula-for-unbalanced-group}
Let a positively unbalanced group~$ \mathcal{C} $ contain~$ N^+ $
(respectively,~$ N^- $) positive (respectively, negative) charges and
$ S^+ $ (respectively,~$ S^- $) positive (respectively, negative) null
points.
Then the following equality is satisfied:
\begin{equation}
\label{eq:eiler-poincare-formula-for-unbalanced-group}
1 + N^- - S^+ + S^- - N^+ = 0.
\end{equation}
\end{lm}

\textbf{Proof.}
Let~$ {\bf v}(\mathcal{C}) $ be the magnetic field
in~$B (\mathcal{C}) $ formed by the group~$ \mathcal{C} $ and
$ {\bf V}(\mathcal{C}) $~be a continuation of
the field~$ {\bf v}(\mathcal{C}) $ to the 3D sphere~$ \mathbb{S}^3 $.
The charges and null points of the magnetic vector
field~$ {\bf v}(\mathcal{C}) $ are the equilibrium states of
the Morse--Smale vector field~$ {\bf V}(\mathcal{C}) $.
We remind that~$ {\bf V}(\mathcal{C}) $, as compared
to~$ {\bf v}(\mathcal{C}) $, has an additional sink whose
topological index is equal to unity.

Morse index (dimensionality of the unstable manifold) of the positive
null point equals unity.
Consequently, the topological index of such a point equals minus unity.
Similarly, the topological index of a negative null point equals unity,
because its Morse index equals two.
Morse index of a negative (respectively, positive) charge equals zero
(respectively, three).
Therefore, the topological index of a negative (respectively, positive)
charge equals unity (respectively, minus unity).
As is known, Euler characteristic of 3D sphere equals zero.
Using the Euler--Poincar{\'e} formula, which states that a sum of topological
indices of the equilibrium states is equal to the Euler characteristic,
we obtain the required result.
$ \Box $

\begin{cor}
\label{cor:from-lm-eiler-poincare-formula-for-unbalanced-group}
Let the conditions of
lemma~\ref{lm:eiler-poincare-formula-for-unbalanced-group} be satisfied.
If there are no negative charges in the sink region ($ N^- = 0 $), then
$$ S^- = (N^+ - 1) + S^+ \geq N^+ - 1. $$
\end{cor}

Let us introduce the partial order~$ \prec $ in the set of equilibrium
states~$ Fix~(f^t) $ of the flow~$ f^t $.
For~$ p, q \in Fix~(f^t) $, let us define that $ p \prec q $ if
$ W^s(p) \cap W^u(q) \neq \emptyset $.
It is convenient to present the above order in the graph whose points are
identified with the equilibrium states~$ Fix~(f^t) $.
The graph vertices corresponding to~$p, q \in Fix~(f^t) $ and related by
the order $ p \prec q $ are connected by the arc directed from~$ q $ to
the point~$ p $.
Such a directed graph~$ \Gamma(f^t) $ is sometimes called
\textit{Smale graph (or diagram)}.

Let us denote a union of all unstable 1D manifolds of the saddles and
all sinks of the flow~$ f^t $ by~$ A(f^t) $.
It is known~\cite{Grines_10} that $ A(f^t) $~is a connected 1D subgraph of
the graph~$ \Gamma(f^t) $, whose vertices are identified with the respective
saddles and sinks.
In this case, arcs of the subgraph correspond to the 1D unstable
separatrices, and they are supplied with the directions from the saddles
to sinks.
Moreover, $A(f^t)$~is the attracting set of the flow~$ f^t $~\cite{Grines_10}.
Similarly, let us denote a union of all stable 1D separatrices of the saddles
and all sources by~$ R(f^t) $.
Then, $ R(f^t) $~is a connected oriented subgraph, which is a repelling set
of the flow~$ f^t $~\cite{Grines_10}.

\begin{figure}
\center{\includegraphics[width=11cm]{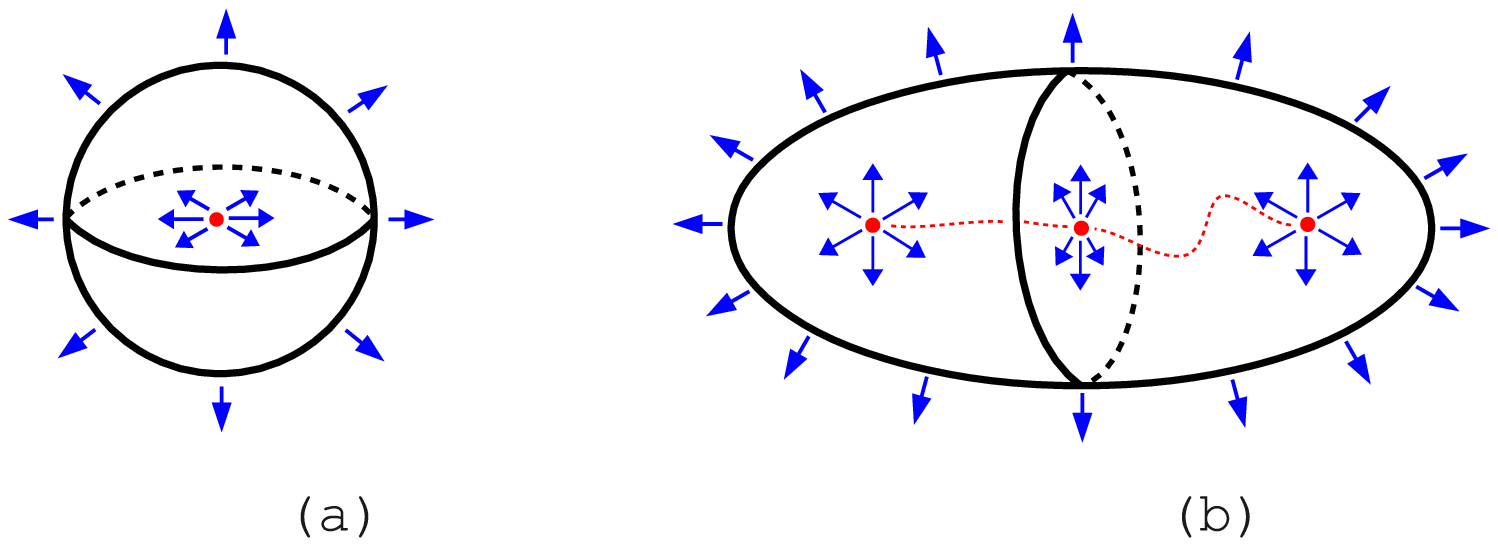}}
\caption{\label{fig:unst-atoms}
Vector field~$ {\bf v}_{\rm sourse} $~(a) and vector
field~$ {\bf v}_{\rm uns} $ of type~(2;~1) in
the ball~$ \mathbb{B}^3 $~(b).}
\end{figure}

To describe a topological structure of the vector fields, we need some
canonical fields.
The vector field~$ {\bf v}_{\rm  sourse} $ will be called
\textit{the source of type~(1;~0)}.
Let us consider the vector field~$ {\bf v}_{\rm uns} $ in
the ball~$ \mathbb{B}^3 $ directed outwards at the ball
boundary~$ \partial\mathbb{B}^3 = S^2 $ and possessing $ l \geq 2 $~sources
$ {\alpha}_1 , \ldots , {\alpha}_l $ and $ l-1 $~saddles
$ {\sigma}_1 , \ldots , {\sigma}_{l-1}$ with Morse index 2.
Such vector field~$ {\bf v}_{\rm uns} $ will be called
\textit{the source of type~(l;~l--1)}.
Structure of the vector field of type~(2;~1) is shown in
Fig.~\ref{fig:unst-atoms}(b).
The vector field~$ {\bf v}_{\rm sourse} $ can be treated as the source
of type~(1;~0), as illustrated in Fig.~\ref{fig:unst-atoms}(a).

Let as assume that
$ {\bf v}_{\rm sink} = -{\bf v}_{\rm sourse} $ and
$ {\bf v}_{\rm stab} = -{\bf v}_{\rm uns} $.
The vector field~$ {\bf v}_{\rm sink} $ is directed inwards at
the ball boundary~$ \partial\mathbb{B}^3 = S^2 $ and has one sink inside
the ball~$ \mathbb{B}^3 $.
The vector field~$ {\bf v}_{\rm stab} $ is directed inwards at
the ball boundary~$ \partial\mathbb{B}^3 = S^2 $ and has $ l \geq 2 $~sinks
$ {\omega}_1 , \ldots , {\omega}_l $ and $ l \! - \! 1 $~saddles with
Morse index~1.
Such vector field~$ {\bf v}_{\rm uns} $ will be called
\textit{the sink of type~(l\,;~l\,--1)}.
Without loss of generality we can assume that the above-mentioned vector
fields are orthogonal to the boundary~$ \partial\mathbb{B}^3 = S^2$ and
are unitary at this boundary.

If boundaries of two copies of the ball~$ \mathbb{B}^3 $ are identified with
each other, then we get a 3-sphere~$ \mathbb{S}^3 $.
If a source of the type~($l\,$;~$l\,$--1) is defined in one copy of the ball,
and the field~$ {\bf v}_{\rm sink} $ is defined in another copy of
the ball, then we get a smooth vector field in~$ \mathbb{S}^3 $, which will
be denoted by~$ {\bf V}_{\rm uns}(l;l\!-\!1)$.
In fact, the following statement follows from the works~\cite{Grines_10}
and~\cite{Pilyugin_75}:

\begin{prop}
\label{prop:top-equiv-on-sphere-with-one-sink}
Let the Morse--Smale vector field\/~$ {\bf V} $ be defined in
the 3D sphere~$ \mathbb{S}^3 $ and its nonwandering set be composed of\/
$ l \geq 2 $~sources, $ l-1 $ saddles of Morse index 2 and one sink.
Then $ {\bf V} $ is topologically equivalent
to~$ {\bf V}_{\rm uns}(l;l-1) $.
\end{prop}

\textbf{Proof.}
Let~$ f^t $ denote the Morse--Smale flow generated by the vector
field~$ {\bf V} $.
Since the number of sources is greater than the number of saddles of Morse
index~2 by unity, then the attractive set $ A(f^t) $ is a segment with sinks
and saddles.
Moreover, the saddles and sinks occur alternatively, and the endpoints of
segment~$ A(f^t) $ are the sources.
Following Lemma~1.1 \cite{Pilyugin_75}, the set~$ A(f^t) $ has a ball
neighborhood that is the source of type~$ (l;l-1) $.
There is a sink beyond this neighborhood.
This leads to the required result.
$\Box$

\section{Proofs of the main results}
\label{sec:Proofs}

\textbf{Proof of Theorem~\ref{thm:at-least-many-nul-points}.}
Let $ {\bf v}(\mathcal{C}) $~be the magnetic field
in~$ B(\mathcal{C}) $ formed by the group~$ \mathcal{C} $ and
$ {\bf V}(\mathcal{C}) $~be a continuation of
the field~$ {\bf v}(\mathcal{C}) $ to the 3D sphere~$ \mathbb{S}^3 $.
Therefore, the field~$ {\bf V}(\mathcal{C})$ has an additional
sink as compared to~$ {\bf v}(\mathcal{C}) $.
We need to prove the inequality~$ S^- \geq l-1 $, where $ S^- $~is
the number of negative null points of the vector
field~$ {\bf V}(\mathcal{C}) $.
Let us prove this inequality by the method of mathematical induction
where the inductive step is done by the number~$ l = N^+ $ of the positive
charges (which is equal to the number of sources of
the field~$ {\bf V}(\mathcal{C}) $).
We remind that $ {\bf V}(\mathcal{C}) $~is the Morse--Smale
vector field, which induces the Morse--Smale flow~$ f^t $ in $ \mathbb{S}^3 $.

Firstly, we show that there exists at least one negative null point for
any~$ l \geq 2$ (this will be simultaneously a proof of the initial step
at~$ l=2 $).
Let $ A = A(f^t)$~be a union of all sinks and unstable (1D) separatrices of
all the saddles of Morse index~1.
Let us assume that the above statement is false.
Then the complement of~$ A $ is a union of the nonintersecting unstable
(3D) manifolds of $ l \geq 2$~sources.
Since a complement of the 1D graph~$ A $ is a connected set,
then we get a contradiction to the connectivity of
the set~$ \mathbb{S}^3{\setminus}A $.

Let us assume that the statement is proved for the number of
sources $ 2, \ldots, l \geq 2 $ and show that it will be true
for~$ l+1 \geq 3 $.
As follows from the previous discussion, there exists at least one
saddle~$ \sigma $ with Morse index~2.
Two (1D) stable separatrices~$ Sep^s_1 $ and $ Sep^s_2$ of
the saddle~$ \sigma $ belong to the unstable manifolds of
the sources~$ {\alpha}_1 $ and $ {\alpha}_2 $, respectively.
Let us consider two cases:
(1)~$ {\alpha}_1 \neq {\alpha}_2$ and
(2)~$ {\alpha}_1 = {\alpha}_2 $.
In the first case, the set
$ {\alpha}_1 \cup Sep^s_1 \cup \sigma \cup Sep^s_2 \cup {\alpha}_2 = S $\/
is a repelling set.
It follows from $ {\alpha}_1 \neq {\alpha}_2 $ that this set~$ S $ has
a neighborhood homeomorphic to a 3-ball, which looks like the source.
Then the original flow can be replaced by the flow with one source instead
of two sources~$ {\alpha}_1 $, $ {\alpha}_2$ and the saddle~$\sigma$.
The resulting flow satisfies the inductive assumption.
Since this flow has exactly one source and one saddle less than before,
we get the required estimate for the original flow.
As follows from the above argumentation, if there exists a saddle with
Morse index~2 for which case~(1) is realized, then the inequality
$ S^- \geq l-1 $ is proved.

In the second case, when $ {\alpha}_1 = {\alpha}_2 $, without loss of
generality we can believe that this case is realized for all the saddles
with Morse index~2.
Then each such saddle is uniquely associated with a source,
$ \sigma \mapsto \alpha = {\alpha}_1 = {\alpha}_2 $.
As a result, we get a stronger inequality $ S^-{\geq}\, l $.

Therefore, the inequality~$ S^- \geq l-1 $ is proved for any group of
charges containing $ l \geq 2 $~positive charges.
We note that if there are no negative charges in the group, then $ N^+ = l $
and $ N^- = 0 $, and consequently the inequality~$ S^- \geq l-1 $ follows from
Corollary~\ref{cor:from-lm-eiler-poincare-formula-for-unbalanced-group}.

If $ S^- = l-1 $, then
formula~(\ref{eq:eiler-poincare-formula-for-unbalanced-group}) leads
to~$ S^+ = 0 $, and consequently all the null points are negative.
Therefore, there are no separators in~$ B(\mathcal{C}) $.

Uniqueness of the topological structure follows immediately from
Proposition~\ref{prop:top-equiv-on-sphere-with-one-sink}.
$ \Box $

\bigskip

\textbf{Proof of Theorem~\ref{thm:at-least-one-separator}.}
Let $ {\bf v}(\mathcal{C}) $~be the vector field in~$ B(\mathcal{C}) $
formed by the group~$ \mathcal{C} $ and $ {\bf V}(\mathcal{C}) $~be
a continuation of the field~$ {\bf v}(\mathcal{C}) $ to the 3D
sphere~$\mathbb{S}^3$.
It is specified that $ {N^+}\!= l $, $ {S^-}\!= l $ and $ {N^-}\!= 0 $.
Then formula~(\ref{eq:eiler-poincare-formula-for-unbalanced-group}) leads to
$ {S^+}\!= 1 $.
Let~$ {\sigma}_0 $ denote the single saddle with topological index minus one.

Let $ f^t $~be the Morse--Smale flow generated by the vector
field~$ {\bf V}(\mathcal{C}) $ in the 3D sphere~$ \mathbb{S}^3 $.
We consider the connected 1D graph~$ R(f^t) $ composed of all 1D stable
manifolds of the saddles and all sources of the flow~$ f^t $.
Let us remind that $ R(f^t) $~is a repelling set of the flow~$ f^t $.

\begin{prop}
\label{prop:nbhd-is-solid-torus}
The graph~$ R(f^t) $ has the neighborhood~$ U(R) $ possessing the following
properties:
\begin{itemize}
\item the boundary~$ \partial U(R) $ of the neighborhood~$U(R)$ is transversal
      to the flow~$ f^t $, and trajectories of the flow leave~$ U(R) $ with
      increasing time;
\item the neighborhood~$ U(R) $ is homeomorphic to a solid torus
      (consequently, the boundary~$ \partial U(R) $ is homeomorphic to
      2D~torus);
\item there exists the saddle~$ \sigma \in U(R) $ (the negative null point)
      whose 2D unstable separatrix~$ W^u(\sigma) $ intersects the
      torus~$ \partial U(R) $ along a closed curve homotopic to the null
      meridian of the torus~$ \partial U(R) $.
\end{itemize}
\end{prop}

\textbf{Proof of Proposition~\ref{prop:nbhd-is-solid-torus}.}
According to our conditions, vertices of the graph~$ R(f^t) $ are composed of
$ l $~saddle and $ l $~source points; so that exactly two arcs enter the each
saddle point, and at least one arc leaves the each source point.
Then $ R(f^t) $~contains the simple cycle~$ C $ of type~(2;~3), which is
supplemented by some (probably, zero) number of segments; each of these
segments contains the equal numbers of source and saddle points (which equal
the number of arcs in the segment).
The cycle~$ C $ has a neighborhood that is homeomorphic to a solid torus; and
the trajectories leave it with increasing time, because~$ C $ possesses
the type~(2;~3).
Without loss of generality we can believe that there is no sink in this
neighborhood (otherwise, the neighborhood can be decreased).
Each of the attached segments has a neighborhood homeomorphic to a ball,
and the trajectories leave this ball with increasing time.
Consequently, there exist the required neighborhood~$ U(R) $ without a sink.
Since $ C $~contains at least one saddle, its 2D unstable separatrix must
intersect~$ \partial U(R) $ along a closed curve homeomorphic to meridian of
the torus.
So, Proposition~\ref{prop:nbhd-is-solid-torus} is proved.
$ \diamondsuit $

\bigskip

Note that the graph~$ A(f^t) $, which is an attracting set of
the flow~$ f^t $, represents a simple cycle composed of
the sink~$ {\omega}_0 $, saddle~$ {\sigma}_0 $ and two its 1D unstable
separatrices.
Consequently, $ A(f^t) $~has the neighborhood~$ U(A) $ homeomorphic to
a solid torus, so that its boundary~$ \partial U(A) $ is transversal to
the flow~$ f^t $, and the trajectories enter~$ U(A) $ with increasing time.
For the sufficiently small neighborhood~$ U(A) $, a 2D stable separatrix
of the saddle~$ {\sigma}_0 $ intersects the 2D torus~$ \partial U(A) $ along
a closed simple curve homotopic to a meridian of the torus~$ \partial U(A) $.
Let this curve be denoted by~$ {\mu}_0 $.

Without loss of generality we can believe that the neighborhoods~$ U(R) $
and~$ U(A) $ do not intersect each other.
Since their union contains all equilibrium states of the flow~$ f^t $,
any positive semitrajectory with the initial point at~$ \partial U(R) $ must
intersect the torus~$ \partial U(A) $.
Consequently, the sphere~$ \mathbb{S}^3 $ can be presented as two solid
tori~$ U(R) $ and~$ U(A) $ where their boundaries~$ \partial U(R) $
and~$ \partial U(A) $ are matched to each other.
Let~$ \vartheta: \partial U(A) \to \partial U(R) $ be such homeomorphism that
$ \mathbb{S}^3 = U(A) {\bigcup}_{\vartheta} U(R) $.
As is known, a gluing of two solid tori results in a 3-sphere only when
a meridian in one boundary torus is matched to the parallel (which may be
rotated a few times along the meridian) in another boundary torus.
Consequently, the image of curve~$ {\mu}_0 $ with respect to~$ \vartheta $
is a closed curve, which intersects any closed curve at~$ \partial U(R) $
homotopic to the null meridian of the torus~$ \partial U(R) $.
Consequently, there exists at least one separator.
$ \Box $

\bigskip

The proof of Proposition~\ref{prop:nbhd-is-solid-torus} immediately leads to
the following statement, which will be used later.

\begin{prop}
\label{prop:for-bif-when many-separators}
Let the premises of Theorem~\ref{thm:at-least-one-separator} be satisfied,
and let $ f^t $~be the Morse--Smale flow generated by the vector
field~$ {\bf V}(\mathcal{C}) $ in the 3D sphere~$ \mathbb{S}^3 $,
which is a continuation of the magnetic field in~$ B(\mathcal{C}) $.
Let us assume that the graph~$ A(f^t) $ is a simple closed curve.
Then there exit at least $ l $~separators in~$ B(\mathcal{C}) $.
\end{prop}

\textbf{Proof.}
In designations of Proposition~\ref{prop:nbhd-is-solid-torus} we get that
the unstable 2D separatrix of each saddle from~$ \partial U(R) $
intersects~$ \partial U(R) $ along a closed curve homotopic to null meridian
of the torus~$\partial U(R)$.
Since $ \partial U(R) $~contains $ l $~saddles, there exist at least
$ l $~separators.
$\diamondsuit$

\bigskip

\textbf{Proof of Theorem~\ref{thm:birth-many-separators}.}
For the sake of simplicity, we construct the required family for~$ l=2 $.
As will be seen from our discussion, such family can be constructed
with an arbitrary number~$ l \geq 2$ of the positive charges.

We consider the Euclidean space~$ \mathbb{R}^3 $ endowed with
the cylindrical coordinates~$ ( \varrho; \varphi; z ) $ and Cartesian
coordinates~$ (x,y,z) $.
Let $U_1$ be the neighbourhood of the origin~$ (0,0,0) $ defined by
$ |x|<0.9 $, $ |y|<0.9 $, $ |z|<1 $, and $ U $ the neighbourhood
defined by $ |x|<\frac{1}{2} $, $ |y|<\frac{1}{2} $, $ |z|<\frac{1}{2} $.
Outside of~$U_1$, the vector field~$\vec{F}_0$ is described by the following
system:
\begin{eqnarray}
\label{for-ro}
\dot{\varrho} & = &
  \varrho\cdot \left[\varrho - 1
  + 4\Psi(\varphi)\right], \\
\label{for-fi}
\dot{\varphi} &=& -\sin (2\varphi), \\
\label{for-z}
\dot{z} &=& z,
\end{eqnarray}
where
$$
\Psi(\varphi) =
\sin^2\left(\frac{\varphi-\frac{\pi}{2}}{2}\right)%
\sin^2\left(\frac{\varphi-\pi}{2}\right)%
\sin^2\left(\frac{\varphi-\frac{3\pi}{2}}{2}\right).
$$

The vector field~$ \vec{F}_0 $ can be extended in the neighbourhood~$ U $
so that $ \vec{F}_0 $ is described by the following system:
$$
\left\{
\begin{array}{ccl}
\dot{x} & = & 1 ,\\
\dot{y} & = & -y , \\
\dot{z} & = & z . \\
\end{array}
\right.
$$
Note that $ {\rm div}\, \vec{F}_0|_{U}\equiv 0 $.

All equilibrium states of $\vec{F}_0$ are in the plane~$ z=0 $,
which is a repeller for the vector field defined by this system.
Since the set of equations can be split into the first two
equations~(\ref{for-ro}), (\ref{for-fi}) and the third
equation~(\ref{for-z}), it is sufficient to consider a phase
portrait in the plane~$ z=0 $.
The equilibrium states in this plane can be only at the rays~$ \varphi = 0 $,
$ \varphi = \frac{\pi}{2} $, $ \varphi = \pi $, and
$ \varphi = \frac{3\pi}{2} $.

\begin{figure}
\center{\includegraphics[width=12cm]{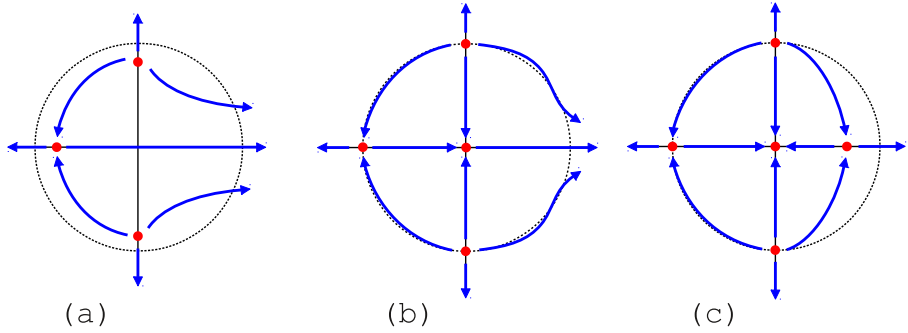}}
\caption{\label{fig:bif-vlad}
Phase portrait in the plane~$ z=0 $.}
\end{figure}

Then, equation~(\ref{for-ro}) takes the form:
$$
\dot{\varrho}=\left\{
\begin{array}{lcl}
\varrho^2 & \mbox{at} & \varphi=0, \; \varrho\neq 0 , \\
\varrho \, (\varrho - 1) &\mbox{at} &
\varphi\in\{\frac{\pi}{2},\pi,\frac{3\pi}{2}\}, \; \varrho\neq 0.
\end{array}
\right.
$$
Consequently, there are three equilibrium states:
$ O_2(1; \frac{\pi}{2})$ is a source,
$ O_3(1; \pi)$ is a saddle, and
$ O_4(1; \frac{3\pi}{2})$ is a source;
see Fig.~\ref{fig:bif-vlad}(a).
The saddle~$ O_3(1; \pi)$ in space has the Morse index~2.
The calculations show that $ {\rm div}\, \vec{F}_0(O_3) = 0 $.

Now, starting with $\vec{F}_0$, we shall construct the family of
vector fields~$ \vec{F}_{t} $ keeping $ \vec{F}_{t} = \vec{F}_0 $ outside
of~$U_1$ and deforming $ \vec{F}_0 $ near the origin $ (0,0,0) $ in
Cartesian coordinates $ (x,y,z) $.
First, the family $ \vec{F}_{\nu}$, $0\leq\nu\leq 1 $, described by
the system
$$
\left\{
\begin{array}{ccl}
\dot{x} & = & (\nu -1)^2+\nu^2x^2 , \\
\dot{y} & = &-y , \\
\dot{z} & = & z(1-2\nu x)\\
\end{array}
\right.
$$ 
moves $\vec{F}_0$ into the vector field $ \vec{F}_{\nu=1} $ defined by
the system
$$
\left\{
\begin{array}{ccl}
\dot{x} & = & x^2 , \\
\dot{y} & = &-y , \\
\dot{z} & = & z(1-2x) . \\
\end{array}\right.
$$
Easy calculations show that $ {\rm div}\, \vec{F}_{\nu} \equiv 0 $ for
any $ 0 \leq \nu \leq 1 $.
So, the saddle--node~$ O_0(0; 0) $ in the origin of coordinates is added to
the previous equilibrium states~$ O_2(0; 1) $, $ O_3(-1; 0) $,
and $ O_4(0; -1) $;
see Fig.~\ref{fig:bif-vlad}(b).
It is convenient to denote $\vec{F}_{\nu=1}$ by $\vec{F}_0$ again.

At last, let us consider the family $ \vec{F}_{\mu}$,
$ 0 \leq \mu \leq \frac{1}{3} $, of vector fields defined by the system:
$$
\left\{
\begin{array}{ccl}
\dot{x} & = & -\mu x+ x^2 , \\
\dot{y} & = &-y , \\
\dot{z} & = & z(1+\mu -2x). \\
\end{array}\right.
$$
Again, it is easy to see that $ {\rm div}\, \vec{F}_{\mu} \equiv 0 $ for
any $ 0 \leq \mu \leq \frac{1}{3} $.
Consequently, the saddle--node~$ O_0(0; 0) $ in the
plane~$ (x; y) $ decays into the sink~$ O_0(0; 0) $ and
saddle~$ O_1(\frac{1}{3}; 0) $;
see Fig.~\ref{fig:bif-vlad}(c).
Therefore, two separators are formed.
$ \Box $

\section{Discussion and Conclusions}
\label{sec:Conclusion}

\begin{enumerate}

\item
Using the Morse--Smale theory, we derived a set of constraints on the number
of the magnetic-field sources (the effective ``magnetic charges'') and
the null points of various types (positive and negative), which should be
a valuable tool for analyzing the structure of complex magnetic fields,
particularly, in the solar anemone flares.
On the one hand, the formulas presented in
Sections~\ref{sec:Summary}--\ref{sec:Proofs} are less powerful than the ones
derived in the old paper~\cite{Gorbachev_88}, because they are based on
the purely topological consideration and, therefore, do not specify any
relations between the positions of the magnetic charges and null points.
On the other hand, these formulas are more general than the previously-known
ones, because they are applicable to the arbitrary number of the charges.
In particular, we present configurations of the charges and null points
of a given type such that no separators exist.
(This does not mean that the given type of the group forbids the existence
of separators.)

\item
An important prerequisite for application of the Morse--Smale inequalities
is the requirement that the group of the magnetic charges is positively
(or negatively) unbalanced, as defined in Section~\ref{sec:Summary}.
Of course, this narrows the scope of applicability of the above-mentioned
inequalities to the particular configurations of solar magnetic fields.
However, as was demonstrated in the recent observational
work~\cite{Dumin_20}, the anemone microflares often develop in the regions
with unbalanced magnetic polarity.
So, the applicability of the Morse--Smale constraints to the these cases
is well justified.

\item
At last, attention should be paid to the correct physical interpretation
of our mathematical constraints.
Namely, as follows from the discussion in Section~\ref{sec:Magnetic_charges},
if $ S^{\pm}_{\rm in} $ is the number of ``physical'' null points in the plane
of magnetic charges and $ S^{\pm}_{\rm out} $ is their number out of this
plane, then
$ S^{\pm} = S^{\pm}_{\rm in} + 2 S^{\pm}_{\rm out} $.
On the other hand, all magnetic charges should be taken with the coefficient
of unity, because in all physically-relevant situations they must be localized
in the same plane.

\end{enumerate}

\section*{Acknowledgements}

YVD is grateful to P.M.~Akhmet'ev, A.T.~Lukashenko, A.V.~Oreshina, and
I.V.~Oreshina for valuable discussions and consultations, as well as
to the Max Planck Institute for the Physics of Complex Systems
(Dresden, Germany) for hospitality during his visits there.
EVZ and VSM were supported by the Laboratory of Dynamical Systems and
Applications of the National Research University Higher School of Economics
(HSE) of the Ministry of Science and Higher Education of RF
[grant ag.\ N$^{\underline{\rm o}}$\,075-15-2019-1931].

We are grateful to W.~Cao, B.~Chen, and P.R.~Goode from the Big Bear
Solar Observatory (BBSO) for the permission to reproduce the middle panel
of Fig.~\ref{fig:observ_data}.
BBSO operation is supported by NJIT and US NSF AGS-1821294 grant.
GST operation is partly supported by the Korea Astronomy and Space
Science Institute, the Seoul National University, and the Key Laboratory
of Solar Activities of Chinese Academy of Sciences (CAS) and
the Operation, Maintenance and Upgrading Fund of CAS for Astronomical
Telescopes and Facility Instruments.

Declarations of interest: none.


\bibliographystyle{elsarticle-num}

\end{document}